\documentstyle[epsf,12pt]{article}

\newcommand{\be}{\begin{equation}}
\newcommand{\ee}{\end{equation}}
\newcommand{\bea}{\begin{eqnarray}}
\newcommand{\eea}{\end{eqnarray}}
\newcommand{\beann}{\begin{eqnarray*}}
\newcommand{\eeann}{\end{eqnarray*}}
\newcommand{\beasn}{\begin{sneqnarray}}
\newcommand{\eeasn}{\end{sneqnarray}}

\catcode`@=11
\def\dif{{\rm d}}
\def\deriv{\@ifnextchar[{\@deriv}{\@deriv[]}}
   \def\@deriv[#1]#2#3{\mathchoice%
{{\dif^{#1}#2\over\dif{#3}^{#1}}}{{\dif^{#1}#2/\dif{#3}^{#1}}}%
{{\dif^{#1}#2\over\dif{#3}^{#1}}}{{\dif^{#1}#2/\dif{#3}^{#1}}}}

\def\presup#1{{}^{#1}\kern-.15em\relax}      
\def\presub#1{{}_{#1}\kern-.12em\relax}      

\def\secteqno{\@addtoreset{equation}{section}%
\def\theequation{\thesection.\arabic{equation}}}
\def\endsecteqno{\def\theequation{\@ifundefined{chapter}%
{\arabic{equation}}{\thechapter.\arabic{equation}}}}
\newcounter{subequation}
\def\thesubequation{\alph{subequation}}
\def\sneqnarray{\stepcounter{equation}\let\@currentlabel=\theequation
\setcounter{subequation}{1}
\def\@eqnnum{{\rm (\theequation\thesubequation)}}
\global\@eqcnt\z@\tabskip\@centering\let\\=\@eqncr\let\@@eqncr=\@@sneqncr
$$\halign to \displaywidth\bgroup\@eqnsel\hskip\@centering
 $\displaystyle\tabskip\z@{##}$&\global\@eqcnt\@ne
 \hskip 2\arraycolsep \hfil${##}$\hfil
 &\global\@eqcnt\tw@ \hskip 2\arraycolsep $\displaystyle\tabskip\z@{##}$\hfil
  \tabskip\@centering&\llap{##}\tabskip\z@\cr}
\def\endsneqnarray{\@@sneqncr\egroup $$\global\@ignoretrue}
\def\@@sneqncr{\let\@tempa\relax
   \ifcase\@eqcnt \def\@tempa{& & &}\or \def\@tempa{& &}
   \else \def\@tempa{&}\fi
     \@tempa \if@eqnsw\@eqnnum\stepcounter{subequation}\fi
     \global\@eqnswtrue\global\@eqcnt\z@\cr}
\def\nobiblabels{\def\@lbibitem[##1]##2{\@bibitem{##2}}}
\catcode`@=12

\secteqno
\begin{document}


\title{{\bf Next-to-leading nonperturbative
      \\ calculation in heavy quarkonium}}

\author{{\sc A.\,Pineda}\\
        \small{\it{Departament d'Estructura i Constituents
               de la Mat\`eria}}\\
        \small{\it{and Institut de F\'\i sica d'Altes Energies.}}\\
        \small{\it{Universitat de Barcelona, Diagonal, 647}}\\
        \small{\it{E-08028 Barcelona, Catalonia, Spain.}}\\
        {\it e-mail:} \small{pineda@ecm.ub.es} }

\date{\today}

\maketitle

\thispagestyle{empty}

\begin{abstract}

The next-to-leading nonperturbative contributions to heavy quarkonium systems
are calculated. The
applicability of the Voloshin-Leutwyler approach to heavy
quarkonia systems for the physical cases of Bottomonium and Charmonium is
investigated.
We study whether the background gluon field correlation time can be considered to be infinity or
 not,
by calculating the leading correction to this assumption and checking whether
the expansion is
under control. A phenomenological analysis of our results is also performed.
The results make us feel optimistic about the
$\Upsilon(1S)$ and to a lesser extent about the $J/\psi$ but do not
about higher levels.
We also briefly discuss the connection with
different models where a finite gluon correlation
time is introduced.


\end{abstract}
\medskip



\vfill
\vbox{
\hfill November 1996\null\par
\hfill UB-ECM-PF 96/17}\null\par

\clearpage



\section{Introduction}
\indent

\bigskip

In the early eighties Voloshin and Leutwyler (VL)
\cite{Vol1,Vol2,Leut}
developed a theory for quark-antiquark bound state systems in the limit
$m_Q \rightarrow \infty$ from first principles. In this approach it was considered that the quarkonia
system can be mainly understood as a Coulomb type
bound state (this is certainly true if the heavy quark mass is large
enough). Moreover, the nonperturbative (NP) contributions are corrections and can be
included
systematically. The leading NP corrections were calculated in refs.
\cite{Vol2,Leut}.
\medskip

Let us briefly discuss the main features of this approach. First,
for calculation convenience, it is usually chosen the
modified
Schwinger gauge fixing for the background gluon field (see \cite{gauge})
\be
x^{i}A^{i}({\vec x},t)=0 \,, \quad  A^0({\vec 0}, t)=0 \,,
\ee
while the Coulomb gauge is used for the perturbative gluon field. 
The Hamiltonian then reads

\be
H=H_{Q {\bar Q}}+H_g+H_I\,,
\ee
where
\be
H_{Q {\bar Q}}=P_sH_s+P_8H_8 \,,
\ee
\be
H_s= -{\Delta \over m} - {C_F {\tilde \alpha}_s(\mu) \over r}\,, \quad
H_8= -{\Delta \over m} + {1 \over 2N_c} {{\tilde \alpha}_s(\mu) \over
r}\,, \ee
$P_s$ ($P_8$) is the singlet (octet) projector on a quark-antiquark
pair,
${\tilde \alpha}_s$ is defined below,
$H_g$ acts on the gluonic and light quark degrees of freedom and
\be
\label{hint}
H_I= - {g \over 2} \xi^a {\vec x} \cdot {\vec E}^a(0) \,,
\ee
where $\xi^a= t^a_1-t^a_2$ ($t^a_1$ ($t^a_2$) is the color $SU(3)$
generator
for the quark (antiquark) with $t^a_{1,2}= {\lambda^a \over 2}$) and
${\vec E}^a$ is the background chromo-electric field.
Giving (\ref{hint})
for $H_I$ we are assuming that the multipole expansion holds. That is
\be
\label{me1}
|{\vec x} | \Lambda_{QCD} <<1
\ee
but we will also assume $H_I$ to be small as far as we will work
within standard quantum mechanics perturbation theory (for
consistency, we will demand the NP corrections to be small)\footnote{If
we did
not consider $H_{int}$ to be small, $|{\vec x}|$ would be an arbitrary
function of $m_{Q}$, $\alpha_{s}$ and $\Lambda_{QCD}$. This more general
situation is beyond the scope of this paper.}. Thus, (\ref{me1}) becomes
\be
\label{me}
{m \beta_n \over n} >> \Lambda_{QCD} \,,
\ee
where $\beta_n \equiv C_{F}{\tilde \alpha_s}/n$.
Nevertheless, for obtaining the standard VL results other assumption is
needed, that is, to
consider the background
gluon field correlation time $T_g$ to be approximately infinity
\be
\label{ae}
{H_g \over m \beta_n^2} << 1 \longrightarrow m \beta_n^2 >> {1 \over
T_g} \sim \Lambda_{QCD} \,. \ee
Now, we have a double expansion $O(({\Lambda_{QCD}n
\over m \beta_n})^2, ({\Lambda_{QCD} \over m \beta_n^2})^2)$.
We remark that (\ref{ae}) is more likely to fail than (\ref{me})
(although, certainly, numerical factors can play a role).
The main
aim of this paper is studying the assumption (\ref{ae}) and its corrections
which, obviously, are going to be the dominant NP corrections.
\medskip

During several years it was generally believed that Bottomonium and
Charmonium systems were not heavy enough. Thus, the
above mention approach could not be applied. Nevertheless, the
situation was not so clear
for the $\Upsilon(1S)$ and for the $n=2$ fine and hyperfine
splitting.
Later on, assumption (\ref{ae}) was relaxed
and $T_g$ was taken into account
\cite{Gromes,Giac1,Dosch} (for a recent study see \cite{SY}).
 A
detailed study of the $\Upsilon(1S)$ and the $n=2$ hyperfine splitting was done in
 the last paper of ref. \cite{Dosch}.
Unfortunately, these approaches are model dependent since they
have to deal with an arbitrary function. Usually, an
 exponential decaying ansatz is used.

    Recently, there have been some attempts to perform a rigorous
    QCD
    determination of the Bottomonium and Charmonium
    properties \cite{Ynd1,Ynd2} using only the VL approach.
It is claimed that a consistent theory for the lower Bottomonium levels
($n=1,2$) and to a lesser extent for the $J/\psi$ is found. Radiative, NP 
and eventually relativistic corrections were put together for the first time.
Some general features for spin independent observables can be inferred
from this study. The leading perturbative and NP corrections are consistently 
smaller than the
Coulomb energy for the $\Upsilon(1S)$ mass. Likewise, the corrections are under control
for the $\Upsilon(1S)$ decay width
 although the result is quite dependent on
the scale. For $n=2$ observables the situation is much
more doubtful since the perturbative and NP corrections are almost as
important as the leading term. Nevertheless, quite
good agreement
with the data was obtained. For these observables and the $\Upsilon(1S)$ decay width the
NP corrections play a fundamental role in order to agree with experiment.
\medskip

It could be argued that further orders in the relativistic, radiative
and NP corrections are needed in order to improve these results. Moreover, so far, no
model independent check of approximation (\ref{ae}) has been done though
it is the most likely to fail. This check is clearly very
important to clarify in which situation we are, that is, whether (\ref{ae})
holds or not and in which observables. In the former case the VL approach
should be applied while in the last case $T_g$ has to be taken into account
in some way and the models displayed in refs. \cite{Gromes}-\cite{SY} may be applied. In order to disentangle this issue,
we calculate the next-to-leading spin independent
NP contributions to the energy levels and wavefunctions of a heavy
$Q{\bar Q}$ pair. We obtain analytical expressions for them, and
we check whether the expansion is
under control.
\medskip

Let us briefly comment upon the fine and hyperfine splitting ($n=1,2$).
For them the agreement with the data (when available) is quite
good according to ref. \cite{Ynd2} (although errors are large).
This
is quite surprising since Dosch and collaborators \cite{Dosch} also find
agreement with the data 
using a model where $T_g$ is taken into account.
Undoubtedly, this should require a detailed study which goes beyond the
scope of this work and
we will not perform it here.
 We expect these observables to depend on shorter distances being
the NP contributions smaller. Combinations of these observables can be
built such that the NP corrections are under control as it was done in
\cite{Ynd3} (see also \cite{yo1} where a omitted contribution was
calculated).
\medskip

%
%


We distribute
the paper as follows. In sec. 2 we calculate the  next-to-leading NP corrections to the energy and
decay width.
In sec. 3 we perform a phenomenological study of our
results.
The last section is devoted to the conclusions
and the discussion of models where $T_g$ is
taken into account.
A few formulas are relegated to Appendix A and we discuss the role
played by the octet potential in Appendix B.
\bigskip

\section{Theoretical results}
\indent
\medskip

We display the calculations and results for the next-to-leading NP
corrections to the spin independent energy levels and wavefunctions.


\subsection{Energy Levels}
\indent

In this subsection we calculate the next-to-leading NP contributions
to the energy levels.
Using the multipole expansion (\ref{me}) and standard Quantum Mechanics
 perturbation
theory the energy correction reads

\bea
\label{deltan}
\nonumber
&\delta E_{nl} &=
\langle 0 \vert
\langle n,l \vert
H_I {1 \over E_n - H_8 -H_g }
H_I \vert n,l \rangle
\vert  0 \rangle
\\
&&
= {g^2 \over 18}
\langle 0 \vert E^{a}_j(0)
\langle n,l \vert
{\vec r}  {1 \over E_n - H_8 -H_g }
{\vec r} \vert n,l \rangle
E^{a}_j (0) \vert  0 \rangle \,.
\eea
$E_n$ is the Coulomb singlet bound state energy.
 $n$, $l$ are the radial and orbital quantum numbers. As far as we
do not study the fine and hyperfine splittings the corrections do not
depend on $j$ (total angular momentum) and $s$ (spin) so we will not
display these indices in the states.
$H_g \sim \Lambda_{QCD}$ and
$H_8 \sim E_n$.

The octet propagator mixes low O($\Lambda_{QCD}$) and high energies
O($E_n$). No assumption about (\ref{ae}) has still been done in
(\ref{deltan}). Let
us assume (\ref{ae}) holds, it follows that an OPE can be performed where the
parameter expansion is of order
\be
\left( {H_g \over E_n - H_8}\right)^2 \sim
\left( {\Lambda_{QCD} \over m \beta_n^2} \right)^2
\ee
and we obtain
\be
\delta E_{nl} = \sum_{r=0}^{\infty} C_r O_r\,,
\ee
where

\be
C_r =
\langle n,l \vert
{\vec r} \left( {1 \over E_n - H_8} \right)^{2r+1}
{\vec r} \vert n,l \rangle
\ee
and
\be
O_r =
{g^2 \over 18}
\langle 0 \vert E^{a}_j
H_g^{2r}
E^{a}_j \vert  0 \rangle\,.
\ee
Using the equation of motion, the gauge fixing and Lorentz
covariance we obtain

\bea
&& O_r =-
{g^2 \over 54}
v^{\beta_0}...v^{\beta_r}v^{\alpha_0}...v^{\alpha_r}
\\
\nonumber
&&
\times
\langle 0 \vert
Tr \left(
[D_{\beta_1}(0),[...[D_{\beta_r}(0),G_{\beta_0 \rho}(0)]...]
[D_{\alpha_1}(0),[...[D_{\alpha_r}(0),{G_{\alpha_0}}^{\rho}(0)]...]
\right) \vert  0 \rangle
\eea

where $v$ is the velocity of the center of masses frame with $v^2=1$
(in the comoving frame $v=(1, {\vec 0})$) and the trace is in the
adjoint representation.

Let us give the order of each term ($ <r> \sim {n \over m \beta_n}$)
\be
\delta E^{(r)}_{nl} \equiv C_r O_r \sim
m \beta_n^2 \left( {\Lambda_{QCD}n \over m \beta_n} \right)^2
\left( {\Lambda_{QCD} \over m \beta_n^2} \right)^{2r+2}
\ee

For $\delta E_{nl}^{(0)}$ we obtain the standard VL result
\cite{Vol2,Leut}.
Nevertheless, we are interested in the next correction
\be
\label{en1}
\delta E^{(1)}_{nl} = C_1 O_1 \sim
m \beta_n^2 { \Lambda_{QCD}^6 n^2 \over m^6 \beta_n^{10}} \,,
\ee
where
\be
\label{c1}
C_1 =
\langle n,l \vert
{\vec r} \left( {1 \over E_n - H_8} \right)^{3}
{\vec r} \vert n,l \rangle\,.
\ee
After a rather lengthy calculation we obtain
\be
C_1= { -1 \over m^5 \beta_1^8} H(n,l)\,,
\ee
where $H(n,l)$ is a dimensionless function which can be obtained using
techniques developed in ref. \cite{Vol2}. We refer the reader to
Appendix A for the analytical expression which can be written as a
rational function. There, we also provide some
numbers for the lower levels. It can be easily checked that it has the
right leading $n^{10}$ dependence.
\medskip

For $O_1$, using Lorentz covariance,
Bianchi identities and the equation of motion, we obtain

\be
\label{o1}
O_1 =
{1 \over 108}\left\{
- g^4 \sum_{A,B} \langle 0| :{\bar q}^A t^a \gamma_{\nu} q^A {\bar q}^B
t^a \gamma^{\nu} q^B: | 0 \rangle +
{3 \over 4} \langle G^3 \rangle  \right\}\,,
\ee
%
where
\be
\langle G^3 \rangle \equiv
g^3 f_{abc}
\langle 0 \vert :
G^{\nu}_{a \mu} G^{\mu}_{b \rho} G^{\rho}_{c \nu} :
\vert  0 \rangle\,.
\ee
and $A,B$ are SU(3) flavor indices.

We can work (\ref{o1}) further using SU(3) flavor symmetry and the
factorization hypothesis. Finally, it reads

\be
O_1 =
{1 \over 108}\left\{ {2^6 \over 3} \pi^2 \alpha_s^2 \langle 0 \vert
{\bar q} q \vert 0 \rangle^2 + {3 \over 4} \langle G^3 \rangle
\right\}\,. \ee

The physical cases will be studied in the next section. Anyway, let us
comment that $O_1$ is going to be positive and,
therefore, the energy correction to be negative.

In Appendix B the error made if we neglect the octet
potential in (\ref{c1}) is discussed.
\medskip


\subsection{Decay Width}
\indent

In this subsection the corrections to the decay width of the
$n^3S_1$ levels of quarkonium are calculated. For further details we refer to
refs. \cite{Vol1,Vol2}.

The decay width of the $n^3S_1$ levels of quarkonium reads
\be
\Gamma (n^3S_1 \rightarrow e^+e^-) =
\pi \left[ {4 \alpha_{em} Q\over M(n^3 S_1)} \right]^2 \left(1-{16\alpha_s(\mu) \over 3\pi
} \right)
 {\
\lower-4.2pt\vbox{\hbox{\rlap{\Large{Res}}\lower9pt\vbox
{\hbox{$\scriptscriptstyle{E=E_{pole}}$}}}}\
}
\langle {\vec x} =0 |G_s(E)|  {\vec y} =0 \rangle\,,
\ee
where $Q$ is the quark charge and
\be
G_s(E)=P_s \langle 0 |{1 \over H-E} |0 \rangle P_s
\ee
is the nonrelativistic propagator (Green function) of the system
projected on the colorless sector of a quark-antiquark pair and the
gluonic vacuum. $E$ is the energy measured from the threshold $2m$. Near
the pole $n$ we have the expansion
\bea
\nonumber
&&\langle {\vec x} =0 |G_s(E)|  {\vec y} =0 \rangle =
{ \rho_n + \delta \rho_n \over E_n +\delta E_{n0} -E} + O((E_n +\delta
E_{n0} -E)^0) \\ && =
{ \rho_n \over E_n -E}- { \rho_n \delta E_{n0} \over (E_n -E)^2}+
{ \delta \rho_n \over E_n -E} + O((E_n -E)^0)+O(\delta E_{n0}^2) \,.
\eea
On the other hand using (\ref{me}) we obtain
\be
\langle {\vec x}=0 \vert
G_s(E)\vert {\vec y}=0 \rangle \simeq \langle {\vec x}=0 \vert
G_s^{(0)}(E)\vert {\vec y}=0
\rangle + \langle {\vec x}=0 \vert \delta G_s(E) \vert {\vec y}=0
\rangle \,,\ee
where
\be
\langle {\vec x}=0 \vert G_s^{(0)}(E) \vert {\vec y}=0 \rangle =
\langle {\vec x}=0 \vert {1 \over H_s-E} \vert {\vec y}=0 \rangle=
{\rho_n \over E_n -E}+O((E_n -E)^0)
\ee
$$
\rho_n = {1 \over \pi} ({m \beta_n \over 2})^3
$$
and
$$
\langle {\vec x}=0 \vert \delta G_s(E) \vert {\vec y}=0 \rangle =
P_s \langle 0 \vert
\langle {\vec x}=0 \vert
{1 \over H_s -E }
H_I  {1 \over H_8 +H_g -E }
H_I
{1 \over H_s -E }
\vert {\vec y}=0 \rangle
\vert 0 \rangle P_s
$$
$$
={g^2 \over 18}
\langle 0 \vert E^{a}_j(0)
\langle {\vec x}=0 \vert
{1 \over H_s -E }
{\vec r}  {1 \over H_8 +H_g -E }
{\vec r}
{1 \over H_s -E }
\vert {\vec y}=0 \rangle
 E^{a}_j(0) \vert 0 \rangle
$$
\be
=- { \rho_n \delta E_{n0} \over (E_n -E)^2}+
{ \delta \rho_n \over E_n -E} + O((E_n -E)^0)\,.
\ee
Proceeding in the same way than in the preceding section we are able to
split low from high energies (OPE).
\be
\langle {\vec x}=0 \vert \delta G_s(E) \vert {\vec y}=0 \rangle =
\sum_{r=0}^{\infty} C_r^{G} O_r\,,
\ee
where
\bea
\nonumber
&C_{r}^G &=
\langle {\vec x}=0 \vert
{1 \over H_s -E }
{\vec r}  \left({1 \over H_8 -E }\right)^{2r+1}
{\vec r}
{1 \over H_s -E }
\vert {\vec y}=0 \rangle
\\
&&
= { A_{-2}^{(r)} \over (E_n -E)^2}+
 { A_{-1}^{(r)} \over (E_n -E)}+
O((E_n -E)^0)
\eea
and $O_r$ is the same than above.
Now, from these expressions we can read off the observables we are
interested in, namely
\be
\label{drho}
\delta \rho_n \equiv \sum_{r=0}^{\infty} \delta \rho_n^{(r)} =
\sum_{r=0}^{\infty} A_{-1}^{(r)} O_r\,,
\ee
\be
\delta E_{n0}={ -1 \over \rho_n} \sum_{r=0}^{\infty} A_{-2}^{(r)}O_r\,.
\ee

This provides a new method to obtain the energy corrections for $l=0$
states. We have used them in order to check our results of the previous
subsection. $\delta \rho_n^{(0)}$ and $\delta E^{(0)}_{n0}$ were already
calculated in ref. \cite{Vol2}. We have used their results to check
ours for this lower order. We are mainly interested in the next correction
 $\delta
\rho_n^{(1)}$ which, after some effort, has been calculated exactly. We write it within the dimensionless quantity
\be
{\delta
\rho_n^{(1)} \over \rho_n} = {O_1 \over m^6 \beta^{10}_1}W(n)\,,
\ee
where the analytical expression and numerical values for $W(n)$ are
displayed in Appendix A. We can see that it has the expected
leading $n^{12}$ dependence.

In Appendix B we give analytical expressions for $\delta \rho_n^{(1)}$
and $\delta E^{(1)}_{n0}$ ($l=0$) with an arbitrary octet potential, $ {1
\over 8} \rightarrow {1 \over N_c^2 -1}$. There, we study the
error made by neglecting the octet potential ($N_c \rightarrow \infty$)
in our results.
\bigskip


\section{Phenomenological analysis}
\indent

In this section we confront our results with the data. We expect our 
corrections
to be small when the theory works and the opposite
could be
a signal that the whole approach breaks down. We will try to see the effect
of the new
contributions on the results found by the authors of ref.
\cite{Ynd1,Ynd2} and we will mainly follow their notation.
\medskip

Let us briefly discuss the set of parameters we are mainly going to use
here.
We make $\Lambda_{QCD}$ running between a range of values
compatible with those
 used by the authors of ref. \cite{Ynd1}.
\be
\label{lambday}
\Lambda_{QCD}^{n_f=3}= 250^{+50}_{-50}\, Mev \,.
\ee
The mean value approximately corresponds
to the value found by Voloshin in ref. \cite{Vol3} using sum rules. The
upper bound corresponds to the standard value of $\Lambda_{QCD}$ found
in
DIS (Deep Inelastic Scattering). The lower bound is closer to the values
proposed by sum rules some years ago. For the two gluon condensate we quote
 the value given in ref. \cite{Ynd1}
\be
\label{g2y}
\langle \alpha G^2 \rangle \simeq 0.042 GeV^4 \,\,.
\ee

We will allow $O_1$ to have some errors. It is not our aim to give rigorous 
errors but to see whether the results are sensitive to sensible variations of 
$O_1$.
For the three gluon condensate we choose the value
\be
\label{g3}
\langle G^3 \rangle \simeq 0.045 \pm 0.009 GeV^6 \,,
\ee
which comes from the dilute instanton gas approximation \cite{Shif} while the errors have been estimated using the value obtained in ref.
\cite{chinos}. 
Similar results are obtained using the three-gluon condensate values given in ref. \cite{sm} (sum rules) or ref. \cite{dilat} (lattice)\footnote{Notice that $<G^3>$ calculated in euclidean space has an opposite sign compared to that in Minkowski space.}.

In the sum rule approach it is usually found the quantity
\be
\label{k}
\kappa \equiv \alpha_s <0|{\bar q} q|0>^2 (\mu) \,,
\ee
which appears to be very weak scale dependent once the anomalous dimension is 
taken into account so it is usually considered to be a constant. In principle 
this expression could be calculated once the quark condensate and 
$\alpha_s$ are known at some scale. However, there has been some controversy
about the error made within the vacuum saturation hypothesis and that may
be the four-quark condensate has been underestimated by around a factor $2$ \cite{Rolf}-\cite{npb}. We just take the average value of these references given in ref. \cite{Pich} 
\be
\label{kappa}
\kappa = 3.8 \pm 2.0 \times 10^{-4} GeV^6\,.
\ee
Nevertheless, instead of Eq. (\ref{k}) we actually have
\be
\alpha_s^2 <0|{\bar q} q|0>^2 (\mu)= \alpha_s (\mu)
 \kappa \,
\ee
in $O_1$. The only remaining scale dependence is in $\alpha_s (\mu)$. We stress that $\alpha_s$ has to be computed at the subtraction point scale, generally, the inverse Bohr radius. This value depends on the physical system one is studying. Therefore, $O_1$ is going to be scale dependent, although weakly.
So 
finally, we obtain using (\ref{g3}), (\ref{kappa}) and their errors (slightly 
larger errors would not change the conclusions)
\be
\label{o1er}
O_1(\mu)= \left( 3.13^{+0.62}_{-0.63} + \alpha_s(\mu) 7.41^{+3.9}_{-3.9} 
\right) \times 10^{-4} GeV^6\,.
\ee
\medskip

Let us finally remark that the range of values given in Eq. (\ref{lambday}) 
does not include those obtained from LEP and $\tau$-decay date 
\cite{Pich}-\cite{8}. We quote the value from reference ref. \cite{Narsm}
\be
\alpha_s (M_Z)= 0.118 \pm 0.003
\ee
which approximately corresponds to
\be
\label{lambdan}
\Lambda_{QCD}^{n_f=3} \simeq 383^{+52}_{-48}\,.
\ee

There has also been some controversy about the value of the gluon condensate 
$< \alpha_s G^2 >$ and several claims of larger values can be found in the 
literature. See \cite{Narsm} and references 
therein where the average value 
\be
\label{g2n} 
< \alpha_s G^2 > = 0.081 GeV^4
\ee
was given excluding the SVZ-like value \cite{Shif}. In fact, 
for many of these studies the input values of $\Lambda_{QCD}$ are 
compatible with those 
given in Eq. 
(\ref{lambdan}). This 
region of parameters was uncovered in ref. \cite{Ynd1,Ynd2}. We have also 
studied these new possibilities although our principal aim will be to compare 
with the results of ref. \cite{Ynd1,Ynd2}\footnote{A complete analysis of 
this new range of parameters, including the static two loop potential and 
the relativistic corrections, is under way \cite{nos4}.}. I remark, in order to be 
consistent, that $<G^3>$ should be also changed since \cite{Shif,dilat}
\be
<G^3> \,\, \propto \, \, < \alpha_s G^2 > \,.
\ee
In our case (changing accordingly the errors)
\be
<G^3> \longrightarrow <G^3> = 0.087^{+0.017}_{-0.018} GeV^6 
\ee
and (using the value of $\kappa$ given in (\ref{kappa}))
\be
\label{o1ern}
O_1(\mu) \longrightarrow O_1(\mu)= \left( 6.04^{+1.18}_{-1.25} + \alpha_s(\mu) 7.41^{+3.9}_{-3.9} 
\right) \times 10^{-4} GeV^6\,.
\ee
Now we have two possible sets of parameters. We choose Eqs. (\ref{lambday}), 
(\ref{g2y}) 
and (\ref{o1er}) to be our first set of parameters or set I, while Eqs. (\ref{lambdan}), 
(\ref{g2n}) and (\ref{o1ern}) are our second set of parameters or set II.
\medskip 

Let us now write the general mass formula for any state
\begin{equation}
\label{mnl}
 M(n,l)
=2m_{b}+A_2(n)+A_3(n,l)+\delta E^{(0)}_{nl}+\delta E^{(1)}_{nl}\,,
\end{equation}
where
$$
A_2(n) = -2m_b {C_f^2 {\tilde \alpha_s}^2(\mu) \over 8n^2}\,,
$$
$$
A_3(n,l) = -2m_b {C_f^2 \beta_0 \alpha_s^2 (\mu) {\tilde \alpha}_s
(\mu) \over 8 \pi n^2 }
\left( \ln \left[ {\mu n \over m_b C_f {\tilde
\alpha_s}(\mu)} \right] + \psi (n+l+1) \right) \,,
$$
\begin{equation}
\delta E^{(0)}_{nl}= m_b {\epsilon_{nl} n^6\pi \langle \alpha_s G^2
\rangle \over (m_b C_f {\tilde \alpha_s}(\mu))^4}\,.
\end{equation}
$\epsilon_{nl}$ were first calculated by Leutwyler in ref. \cite{Leut}.
For the levels we are interested in they read
\be
\epsilon_{10}= {1872 \over 1275}\,, \quad
\epsilon_{20}= {2102 \over 1326}\,, \quad
\epsilon_{21}= {9929 \over 9945}\,.
\ee
$\alpha_s$ is
the two-loop running coupling constant. $A_3$, which include
radiative corrections, and
\be
{\tilde \alpha_s} (\mu) = \left[1 +  (a_1 + \gamma_E \beta_0/2){\alpha_s (\mu)
 \over \pi} \right] \alpha_s (\mu)
\ee
$$
a_1= {31C_A -20T_F n_f \over 36}
$$
have been quoted from ref. \cite{Ynd1}. We remark that the relativistic 
corrections have not been included in this analysis. This is due to the 
incomplete knowledge of the $O(m \alpha_s^4)$ corrections for the mass (or 
the equivalent for the decay width) in the $\overline{MS}$ 
scheme because the static two loop potential has not been calculated\footnote{
Recently, we became aware that the static two loop potential had been finally 
calculated \cite{2lpot}. We expect to introduce those results in the near 
future \cite{nos4}.}. Although naively, these corrections are expected to 
be small they 
could blow up if $\alpha_s$, ${\tilde \alpha_s}$ are big enough. In order to 
make quantitative this statement the next-to-next-to-leading perturbative 
contribution, 
including radiative and relativistic corrections, should be calculated. 
Moreover, it has been seen, according to the recent sum rule derivation of 
the Balmer mass formula \cite{Narsm}, that the relativistic corrections tends 
to compensate the radiative coulombic corrections.

Let us briefly comment upon some general features of the numerical study. 
Large values of $\alpha_s$, ${\tilde \alpha_s}$ appear in the problem. 
Moreover, for larger values of $\Lambda_{QCD}$ the value of $\alpha_s$ also 
increases. Therefore, for the second set of parameters, the results are going 
to be more doubtful. This is especially so for $n=2$ Bottomonium 
states and Charmonium. Nevertheless, without the knowledge of the 
next-to-next-to-leading 
perturbative contribution, we do not have a real check in order 
to know when the $\alpha_s$ perturbative expansion breaks down. Therefore, 
we refrain from displaying numbers for $n=2$ Bottomonium states and Charmonium with 
the second set of parameters here (although we do not rule out the possibility 
the VL approach to work for them). This is relegated to future work where the 
static two loop potential and the relativistic corrections 
will be taken into account\cite{nos4} .
\medskip


\subsection{$\Upsilon (1S)$ mass}
\indent

We start our study with the $\Upsilon(1S)$ mass where the formalism is
expected to apply better. We proceed as follows. First of all, we fix
 $m_b$ and the inverse Bohr radius $a_{bb,1}^{-1}$ from the
self-consistency equation
\begin{equation}
 a_{bb,1}^{-1} = {m_b C_f {\tilde \alpha_s}(a_{bb,1}^{-1}) \over 2}\,
\end{equation}
and the $\Upsilon (1S)$ mass set at the inverse Bohr radius scale
($\mu \rightarrow a_{bb,1}^{-1}$).
We allow for the both above mentioned sets of parameters (without using 
errors for $O_1$)
and give the relative
weight of each contribution in table I. It can be seen 
that $\Lambda_{QCD}$ turns out to be the main source of error. We also get
that perturbation theory
works nicely here. Similar results than in ref.
\cite{Ynd1} are found being not to much affected by the new contribution.

%
%

\begin{table}
\begin{center}
\begin{tabular}{|c|c|c|c|c|c|c|}  \hline
$\Lambda_{QCD}^{n_f=3}$
              & $A_2$  & $A_3$
              & $\delta E^{(0)}_{10}$
              & $\delta E^{(1)}_{10}$
              & $m_b$ & $a_{bb,1}^{-1}$     \\ \hline

$250^{+50}_{-50} $        & $-376^{-64}_{+62}$  & $61^{+13}_{-12}$    
              & $18^{-5}_{+7}$ & $-4^{+2}_{-3}$
              & $4881^{+26}_{-27}$  & $1355^{+114}_{-121}$        \\ \hline

$383^{+52}_{-48}$         & $-535^{-74}_{+68}$  & $93^{+16}_{-15}$    
              & $17^{-4}_{+5}$ & $-1^{+0}_{-1}$
              & $4943^{+31}_{-28}$  & $1626^{+115}_{-111}$        \\ \hline
\end{tabular}
\end{center}
\caption{We display $A_2(1,0)$, $A_3(1,0)$, $\delta E^{(0)}_{10}$ and
$\delta E^{(1)}_{10}$ for the $\Upsilon (1S)$.
The last two columns give our
results for $m_b$ and $a_{bb,1}^{-1}$. All the quantities are displayed in 
units of MeV. We have used $\Lambda_{QCD}^{n_f=4}= 330^{+52}_{-48} \, MeV$ for 
the second set of parameters.}
\end{table}

%
%

For the first set of parameters, plugging the errors given in Eq. (\ref{o1er})
 into $\delta E^{(1)}_{10}$ and 
making the ratio
with
$\delta E^{(0)}_{10}$, we find
\be
0.19 <|{\delta E^{(1)}_{10} \over \delta E^{(0)}_{10} }|< 0.38 \,, \quad
0.14 <|{\delta E^{(1)}_{10} \over \delta E^{(0)}_{10} }|< 0.27 \,, \quad
0.10 <|{\delta E^{(1)}_{10} \over \delta E^{(0)}_{10} }|< 0.20 \,,
\ee
for $\Lambda_{QCD}^{n_f=3}= 200$, $250$, $300 \, MeV$ respectively. Likewise, 
for the second set of parameters, we obtain
\be
0.08 <|{\delta E^{(1)}_{10} \over \delta E^{(0)}_{10} }|< 0.14 \,, \quad
0.06 <|{\delta E^{(1)}_{10} \over \delta E^{(0)}_{10} }|< 0.11 \,, \quad
0.05 <|{\delta E^{(1)}_{10} \over \delta E^{(0)}_{10} }|< 0.09 \,,
\ee
for $\Lambda_{QCD}^{n_f=4} = 282$, $330$, $382 \, MeV$ respectively. 
We see $\delta E^{(1)}_{10}$ is always reasonably smaller than $\delta
E^{(0)}_{10}$ and we can safely conclude we are inside the VL regime for
this observable.

Let us comment on some features of the results. They are very stable under
variations of $\mu$ because the $\mu$ scale dependence of $A_2$ and $A_3$
cancel
each other. This can be seen in more detail in Figure I where we plot
the $\Upsilon (1S)$ mass as a function of the
scale. Moreover,
the scale of minimum sensitivity
and the inverse Bohr radius scale are quite close. Indeed, the difference
between the $M_{\Upsilon(1S)}$ values for those two scales is
completely negligible. Other very important point is that not only the NP
corrections are small but also under control. That is, higher
order NP corrections can be calculated giving smaller contributions. The values we obtain for $m_b$ are compatible with the lattice calculation \cite{mb} although somewhat larger than those obtained in sum rules \cite{Narison} if the 
comparison is done with the two-loop perturbative pole mass. Nevertheless, the agreement is quite good if the comparison is done with the three-loop 
pole mass.
\medskip

\medskip
\begin{figure}
\hspace{1.0in}
\epsfxsize=3.8in
\centerline{\epsffile{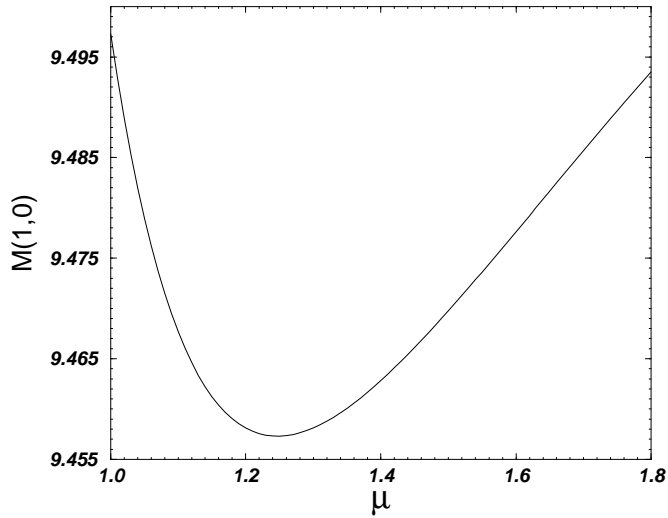}}
\caption {Plot of $M_{\Upsilon(1S)}$ ($GeV$) versus $\mu$ ($GeV$). The NP
corrections have a little effect on the draw. We have used the first set of 
parameters with $\Lambda^{n_f=3}_{QCD} =250 MeV$ and $m_b= 4881 MeV$.
The same features hold for the other values of $\Lambda^{n_f=3}_{QCD}$.}
\label{fig1}
\end{figure}

Let us finally point out one prediction which follows directly from our results, ${\bar \Lambda}$, the nonperturbative parameter relating the mass of the heavy meson ($B$, $D$) to $m_Q$ ($m_b$, $m_c$) in the HQET \cite{Neubert}. We obtain
\be
{\bar \Lambda}=433^{-27}_{+27} MeV \, (\Lambda_{QCD}^{n_f=3}=250^{+50}_{-50} 
MeV) \,; 
\quad  
{\bar \Lambda}=370^{-31}_{+28} MeV \, (\Lambda_{QCD}^{n_f=4}= 
330^{+52}_{-48} MeV)\,,
\ee
for the first and the second set of parameters respectively. Similar 
comments than for $m_b$ apply here when comparing with the literature.
\medskip


\subsection{$n=2$ states}
\indent

We now discuss $n=2$ states and their masses $M(2,0)$ and $M(2,1)$.  We will 
only use the first set of parameters in this subsection. 
We expect the theory to work worse or even to fail completely here. In
fact,
there are several signals that for $n=2$ the theory does not really work.
For instance, we can now predict the physical masses since we have got
$m_b$ and, in order
to set the scale, we can use the self-consistency equation for $n=2$,
that is
\begin{equation}
 a_{bb,2}^{-1} = {m_b C_f {\tilde \alpha_s}(a_{bb,2}^{-1}) \over 4}\,.
\end{equation}
%
%
\begin{table}
\begin{center}
\begin{tabular}{|c|c|c|c|c|c|c|}  \hline
$\Lambda_{QCD}^{n_f=3}$
              & $A_2$  & $A_3$
              & $\delta E^{(0)}_{20}$
              & $\delta E^{(1)}_{20}$
              & $M(2,0)$ & $a_{bb,2}^{-1}$ \\ \hline
$250^{+50}_{-50} $        & $-178^{-37}_{+35}$  & $-29^{-8}_{+7}$    
              & $339^{-109}_{+190}$ & $-439^{+228}_{-586}$
              & $9454^{+128}_{-408}$  & $932^{+095}_{-100}$        \\ \hline
\end{tabular}
\end{center}
\caption{We display $A_2(2,0)$, $A_3(2,0)$, $\delta E^{(0)}_{20}$ and
$\delta E^{(1)}_{20}$.
The last two columns give our
results for $M(2,0)$ and $a_{bb,2}^{-1}$. All the quantities are displayed in MeV. The experimental value is
$M(2,0)=10023 MeV$.}
\end{table}
%
%
%
%
\begin{table}
\begin{center}
\begin{tabular}{|c|c|c|c|c|c|}  \hline
$\Lambda_{QCD}^{n_f=3}$
              & $A_2$  & $A_3$
              & $\delta E^{(0)}_{21}$
              & $\delta E^{(1)}_{21}$
              & $M(2,1)$     \\ \hline
$250^{+50}_{-50} $        & $-178^{-37}_{+35}$  & $-71^{-19}_{+17}$    
              & $213^{-068}_{+120}$ & $-247^{+128}_{-329}$
              & $9479^{+057}_{-211}$    \\ \hline
\end{tabular}
\end{center}
\caption{We display $A_2(2,1)$, $A_3(2,1)$, $\delta E^{(0)}_{21}$ and
$\delta E^{(1)}_{21}$.
The last column gives our
results for $M(2,1)$. All the quantities are displayed in MeV. The experimental value, averaged over angular
momentum, is $M(2,1)=9900 MeV$.}
\end{table}
%
%
We display the results in table II and III. We find out that $|\delta
E^{(0)}_{nl}| >|A_2|$ and also $|\delta E^{(1)}_{nl}| >|\delta E^{(0)}_{nl}|$,
excepting the case\footnote{Nevertheless the reliability of 
perturbation theory is marginal since $|\delta E^{(1)}_{21}|
 {\ \lower-1.2pt\vbox{\hbox{\rlap{$<$}\lower5pt\vbox{\hbox{$\sim$}}}}\ }
  |\delta E^{(0)}_{21}|$ and 
$|\delta E^{(0)}_{21}|
 {\ \lower-1.2pt\vbox{\hbox{\rlap{$>$}\lower5pt\vbox{\hbox{$\sim$}}}}\ }
 {2 \over 3} |A_2|$. Moreover,  
 $\alpha_s = 0.403$ and
${\tilde \alpha_s} = 0.628$ are
quite big making the $\alpha_s$ expansion rather doubtful.
Therefore, knowledge of perturbative and NP next order contributions
would be welcome in order to discern whether the $\alpha_s$ and/or the
NP expansion work or not.}
$n=2,l=1$ for $\Lambda^{n_f=3}_{QCD}=300 MeV$,
so the theory is not really trustworthy. Agreement with the data is also
very bad. We have also studied the dependence on the scale. We
find that the scale we reach the minimum and the scale
we obtain doing the self consistency equation (with $n=2$) are not
so near now, especially for the $n=2$, $l=1$ state.
Therefore, we rather try with mass shifts like
\be
\Delta M(2,0) = M(2,0)-M(1,0) \,, \quad \Delta M(2,1) =M(2,1)-M(1,0) \,,
\ee
as it was indeed done in ref. \cite{Ynd1}.
Let us briefly comment upon this work. The
results
found there were already quite delicate because the corrections were
around
to be as large as the leading term. However, this fact was compensated with the
good agreement with the data obtained.
\be
\label{dmexp}
\Delta M(2,0) = 479 MeV\, (expt.\,\, 563 MeV)\,, \quad \Delta M(2,1) =
417 MeV\, (expt.\,\, 450 MeV) \,.
\ee

From this point of view we should say that
the new contribution completely destroys this agreement as we can see if
we simply plug our result with their input set of parameters
($m_b=4906MeV$, $\mu=986 MeV$ for $\Delta M(2,0)$ and $\mu=1062 MeV$
for $\Delta M(2,1)$) and with the errors given by (\ref{o1er}).

\be
\Delta M(2,0) = -182^{+231}_{-232} MeV\,, \quad \Delta M(2,1) =
-211^{+276}_{-218} MeV\,. \ee
The new contribution even changes the sign of $\Delta M$! This result is
due to $|\delta E^{(1)}|>|\delta E^{(0)}|$.
It can also be seen that the dependence on the scale is very strong. Let
us briefly explain what is going on here. Let us first neglect $\delta
E^{(1)}$. Then, for consistency of the theory, one demands

\be
\label{des}
|A_2|>|A_3| \, \quad |A_2|> |\delta E^{(0)}| \,.
\ee
The first constraint does not allow to lower the scale while the second one
does not allow to increase it, hence, only a very small window is permitted.
(\ref{dmexp}) was found for a
$\mu$ scale such as radiative and NP contributions were equal in
absolute value and even in this optimum situation (\ref{des}) was
not well satisfied ($|A_2| \sim |A_3| \sim |\delta
E^{(0)}|$). Now, if we introduce the new contribution a new
consistency constraint $|\delta E^{(0)}| > |\delta
E^{(1)}|$ arises which is also not well satisfied or, even more frequently, it
is not satisfied at all.

We do not give more numbers here as far as we conclude the theory is not
really
trustworthy for $n=2$ states. Indeed, all these results make us feel
that probably both approximations (\ref{me}) and (\ref{ae}) fail. In
fact, that can be seen by the large mass gap between the $n=1$ and
$n=2$ states which can not be obtained by only perturbing about the Coulomb
spectrum. Therefore, $n=2$ states seem to behave completely
different than $n=1$ states. For the former the NP
contributions are large and can not be treated as corrections so other
approaches should be attempted. For example, a hopeful approach has been
developed by Dosch and Simonov \cite{DS} where they seem to be able to
connect the VL regime with the regime where potential models work.
\medskip




\subsection{Charmonium}
\indent

We now briefly discuss the Charmonium case. We will only use the first set 
of parameters. For this particle the Bohr
scale is rather small and consequently $\alpha_s$ quite large (indeed larger
than for $n=2$ Bottomonium states) so the
$\alpha_s$ perturbative expansion should be taken with care. Let us anyway say a few
words about it. First let us find the value of our contribution with the
set of parameters given in ref. \cite{Ynd1} ($m_c=1570 MeV$ and
$\mu=1GeV$). We obtain (using the errors given in (\ref{o1er}))
\be
\delta E^{(1)}_{10}= -154^{+54}_{-54} MeV\,,
\ee
$|\delta E^{(1)}_{10}|
 {\ \lower-1.2pt\vbox{\hbox{\rlap{$<$}\lower5pt\vbox{\hbox{$\sim$}}}}\ }
 |\delta E^{(0)}_{10}|$ and
$|A_2|=
|\delta E^{(0)}_{10}|$ (this was the condition used to fix the scale in ref.
\cite{Ynd1}) for those parameters so the results are a little
bit doubtful.
Let us now use our standard procedure by fitting the scale with the
self-consistency Bohr radius equation (notice that we use $\eta_c$
rather than $J/\psi$ in our procedure). We display the results in table
IV. Everything seems to behave
quite properly, especially the NP contributions, but for those scales
$\alpha_s$ is quite big so it would be desirable to obtain the
next-to-next-to-leading order correction $A_4$ in order to discern whether the 
$\alpha_s$ expansion expansion can be applied or not.
We notice here that $\delta E^{(1)}_{10}$ and $\delta E^{(0)}_{10}$
behave quite well now, say, the inequalities $|A_2| >|\delta E^{(0)}_{10}| >
|\delta E^{(1)}_{10}|$ are well satisfied. The reason being that
$n=1$ and the blowing up of the NP corrections with $n$ does not arise.

Just as in the $m_b$ case the values we obtain for $m_c$ are somewhat larger 
than those obtained in sum rules \cite{Narison} if the 
comparison is done with the two-loop perturbative pole mass. Nevertheless, the agreement is quite good if the comparison is done with the three-loop 
pole mass.

\medskip

%
%

\begin{table}
\begin{center}
\begin{tabular}{|c|c|c|c|c|c|c|}  \hline
$\Lambda_{QCD}^{n_f=3}$
              & $A_2$  & $A_3$
              & $\delta E^{(0)}_{10}$
              & $\delta E^{(1)}_{10}$
              & $m_c$ & $a_{cc,1}^{-1}$ \\ \hline
$250^{+50}_{-50} $        & $-368^{-85}_{+80}$  & $78^{+23}_{-20}$    
              & $55^{-19}_{+38}$ & $-14^{+08}_{-23}$
              & $1614^{+37}_{-38}$ & $770^{+95}_{-97}$    \\ \hline
\end{tabular}
\end{center}
\caption{We display $A_2(1,0)$, $A_3(1,0)$, $\delta E^{(0)}_{10}$ and
$\delta E^{(1)}_{10}$ for the $\eta_c$.
The last two columns give our
results for $m_c$ and $a_{cc,1}^{-1}$. All the quantities are displayed in MeV.}
\end{table}

%
%


\subsection{$\Upsilon(1S)$ decay width}
\indent

Let us conclude with the $\Upsilon(1S)$ decay width.
\be
\label{gy1}
\Gamma (\Upsilon(1S)) = \Gamma^{(0)} \left( 1 + \delta_r \right) 
\left( 1 + \delta_{WF} + {\delta \rho^{(0)}_1 \over 2 \rho_1}+ {\delta \rho^{(1)}_1 \over 2 \rho_1} \right)^2 \,,
\ee
where 
$$
\Gamma^{(0)} = 2 \left[ {Q_b \alpha_{em} \over M_{\Upsilon(1S)}} \right]^2
(m_b C_f {\tilde \alpha}_s(\mu))^3 \,, \quad
\delta_{WF} = 3 \beta_0
\left( ln \left(
{\mu \over m_b C_f
{\tilde \alpha}_s (\mu)} \right)- \gamma_E \right)
{\alpha_s (\mu) \over 4 \pi} \,,
$$
\be
\delta_r = -{4 C_f \alpha_s (\mu) \over \pi} \,, \quad
{\delta \rho^{(0)}_{1} \over 2 \rho_1}=
\left( {270459 \over 217600}
+ {1838781 \over5780000} \right) {\pi < \alpha_s
G^2 > \over m_b^4 {\tilde \alpha}_s^6 (\mu) }\,.
\ee

We have quoted (\ref{gy1}) from ref. \cite{Ynd1} adding our result. Let us 
first discuss the set I of physical inputs. We have drawn
$\Gamma$ with and without $\delta \rho^{(1)}$ in Figure 2. For small
values of $\mu$ the results are not reliable because the
perturbative corrections become too large. It can also be seen the
strong
dependence on the scale. The results on Figure 2 should be compared with
the experimental value $\Gamma(\Upsilon (1S))=1.34 KeV$	
and with $\Gamma(\Upsilon (1S))=1.12 KeV$, the result found in ref.
\cite{Ynd2} using a value of $\mu
\simeq 2.33 GeV$ ($n_f=4$)
such that the perturbative and NP corrections cancel each other. We
now include our result for such a scale. We
find (using the errors given in (\ref{o1er}))
\be
\Gamma(\Upsilon (1S))=0.36^{+0.20}_{-0.15} KeV \,.
\ee
We see that it destroys agreement with the data. What is going
on here is that for such a scale $|{\delta \rho^{(1)}_1 \over 2 \rho_1}|
{\ \lower-1.2pt\vbox{\hbox{\rlap{$<$}\lower5pt\vbox{\hbox{$\sim$}}}}\ }
1$ and
$|\delta \rho^{(1)}_1| > |\delta \rho^{(0)}_1|$
so the result is not reliable. Indeed, one can not get agreement
with the data for any scale as it
can be seen in Figure 2. If one lowers the scale in order to fulfill
$|\delta \rho^{(0)}_1| > |\delta \rho^{(1)}_1|$ the perturbative
contributions grow in size with negative sign and make agreement
with the data worse.

\medskip
\begin{figure}
\hspace{1.0in}
\epsfxsize=4.2in
\centerline{\epsffile{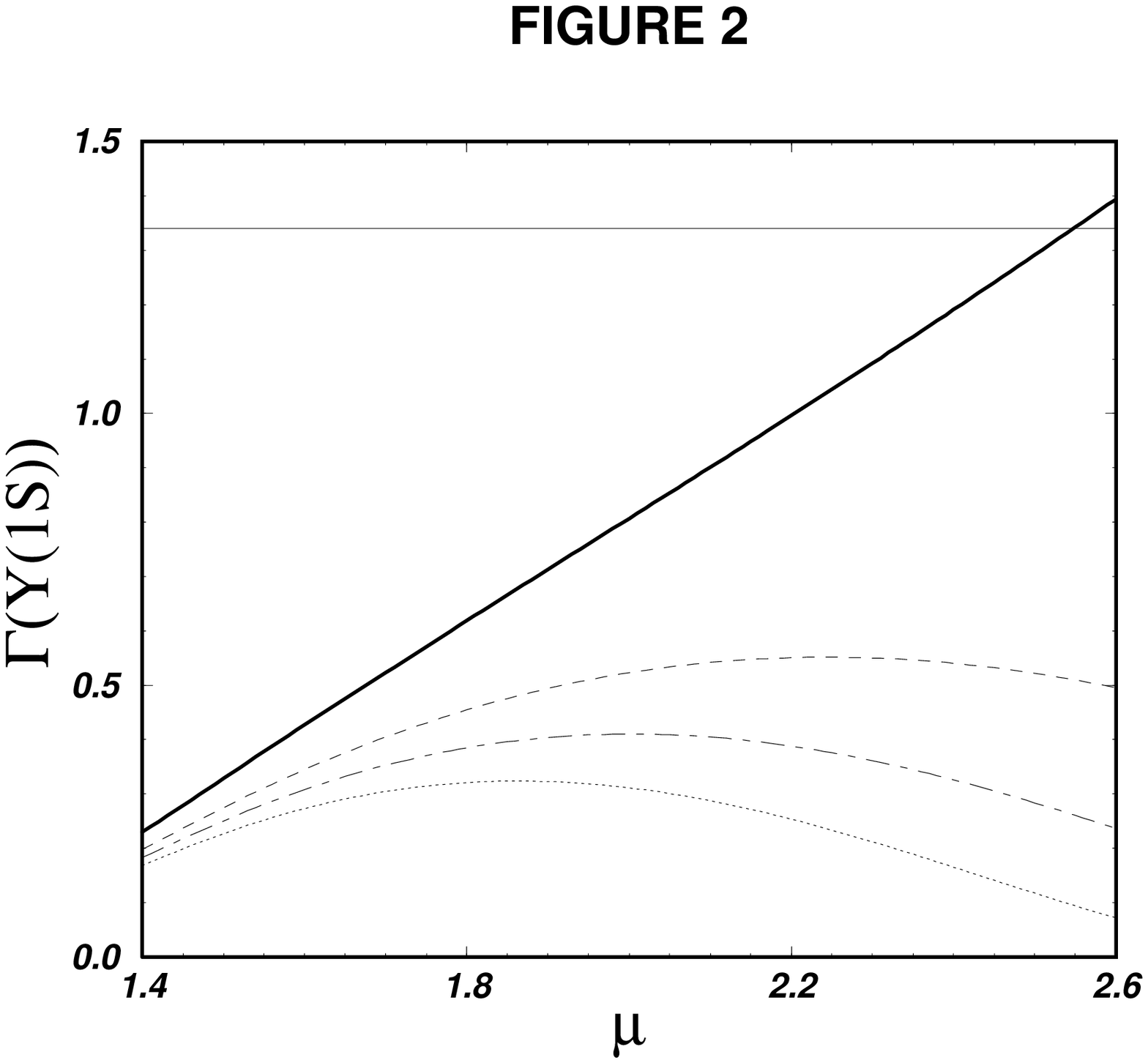}}
\caption{Plot of the $\Upsilon(1S)$ decay width ($KeV$) versus $\mu$ 
($GeV$). We have used the first set of parameters with $\Lambda^{n_f=3}_{QCD} 
=250 MeV$ and $m_b= 4881 MeV$. The continuous line draws the decay width with 
$O_1=0 \,GeV^6$, the dot-dashed line with the mean value displayed in 
(\ref{o1er}), the dashed and dotted line, 
respectively with the lower and
upper value displayed in (\ref{o1er}). The same features hold for the other 
values of $\Lambda^{n_f=3}_{QCD}$. We also plot
the experimental value $\Gamma(\Upsilon (1S))=1.34 KeV$.}
\label{fig2}
\end{figure}

These results are somehow surprising since we would expect to obtain the same features as those obtained in the $M_{\Upsilon(1S)}$ case. One point is that here 
we are faced with a two scale problem, the annihilation scale $O(m_b)$ and the
wavefunction scale $O(a_{bb,1}^{-1})$\footnote{The $M_{\Upsilon(1S)}$ does not have this problem because it is an asymptotic feature of the physical state and only the typical bound state scale, $a_{bb,1}^{-1}$, can appear in the dynamics.}. For this kind of observables factorization holds, that is, we can split the observable in two pieces the
behavior of which is dictated by long and short distances respectively. Effective theories are especially useful in these situations. In our case the suitable effective theory is NRQCD \cite{NRQCD}
which has proved to be extremely successful in separating low from high energies.

Using NRQCD it can be seen that the right scale for $\delta_r$ is $\mu \sim
m_b$. The NP
contributions behave properly for their natural scales $\mu \sim a_{bb,1}^{-1}$, that is,
the inequalities $|\rho_1| >
|\delta \rho^{(0)}_1| > |\delta \rho^{(1)}_1|$ are well satisfied, especially
the first one. The typical scale of $\Gamma^{(0)}$ and $\delta_{WF}$ is also $\mu \sim a_{bb,1}^{-1}$. The problem is that $\delta_{WF}$ is large and strongly dependent on the scale. For instance, we obtain (with $\Lambda^{n_f=3}_{QCD} =250 MeV$, $m_b= 4881 MeV$ and $a_{bb,1}^{-1}=1355 MeV$)
$
\Gamma(\Upsilon(1S)) = 0.11 KeV
$
for $\Gamma^{(0)} (a_{bb,1}^{-1}) = 2.90 KeV$,
$$
\delta_{WF}(a_{bb,1}^{-1})= -0.81 \,, \quad \delta_{r}(m_b)= -0.31 \,, \quad 
{\delta \rho^{(0)}_1 \over 2\rho_1}(a_{bb,1}^{-1})=0.07 \,, \quad
{\delta \rho^{(1)}_1 \over 2\rho_1}(a_{bb,1}^{-1})=-0.02 \,,
$$
while for $\delta_{WF}(2 a_{bb,1}^{-1})= -0.08$ and the remaining set of 
corrections kept fixed we obtain 
$
\Gamma(\Upsilon(1S)) = 1.84 KeV
$. The experimental value lies in between for the value $\delta_{WF}(2.21 GeV)= -0.23$. Although $\delta_{WF}$ is rather large it is not senseless to suppose that higher orders could improve the result changing significantly the value of $\delta_{WF}$.
\medskip

For the second set of parameters the same general features are observed. The 
NP contributions are under control (($|\rho_1| > |\delta \rho^{(0)}_1| > 
|\delta \rho^{(1)}_1|$) but the $\alpha_s$ perturbative expansion behavior is 
even worse. For instance, we obtain (with $\Lambda^{n_f=4}_{QCD} =330 MeV$, $m_b= 4943 MeV$ and $a_{bb,1}^{-1}=1626 MeV$)
$
\Gamma(\Upsilon(1S)) = 0.09 KeV
$
for $\Gamma^{(0)} (a_{bb,1}^{-1}) = 5.02 KeV$,
$$
\delta_{WF}(a_{bb,1}^{-1})= -0.87 \,, \quad \delta_{r}(m_b)= -0.36 \,, \quad 
{\delta \rho^{(0)}_1 \over 2\rho_1}(a_{bb,1}^{-1})=0.05 \,, \quad
{\delta \rho^{(1)}_1 \over 2\rho_1}(a_{bb,1}^{-1})=-0.01 \,.
$$
Summarizing we may conclude that the $\Gamma (\Upsilon(1S))$ belongs to the VL
regime since for scales $\mu \sim a_{bb,1}^{-1}$ the NP contributions are under control ($|\rho_1| > |\delta \rho^{(0)}_1| > |\delta \rho^{(1)}_1|$). However, no precise determination of  $\Gamma (\Upsilon(1S))$ can be given due to the
strong dependence on the scale and the large perturbative corrections 
involved. The solution could be to calculate the
two loop ${\bar Q} - Q$ potential and introduce the next order perturbative 
correction. This may significantly improve the value of $\delta_{WF}$.
\medskip

From the whole set of results we can conclude that the $\Upsilon(1S)$ belongs to the VL regime. Therefore, $T_g$ is not needed and the approach developed in ref. \cite{Dosch} or potential models should not be applied here.
\section{Discussion and Conclusions}
\indent

Let us briefly comment upon models where $T_g$ is fully taken into
account
\cite{Dosch,SY}. There, it is obtained
\be
\label{tg}
\delta E_{nl} =
{ \pi <\alpha_s G^2> \over 18}
\langle n,l \vert
r_i {1 \over H_8 - E_n + 1/T_g}
r_i \vert n,l \rangle\,.
\ee
In order to compare it with our results we should take
the limit $T_g \rightarrow \infty$ in (\ref{tg}). That is, $1/T_g << m
\beta_n^2$. We see that it has
the right asymptotic limit, but it fails to
describe properly the preasymptotic corrections. For
instance, instead of $\delta E^{(1)} \sim \Lambda_{QCD}^6$, (\ref{tg}) already
has $\Lambda_{QCD}^5$ corrections. This is not so strange if we recall
the exponential approximation for the gluon correlator (Euclidean space)
which has been used to obtain (\ref{tg})
\be
\label{lin}
<G(t)G(0)> \sim e^{-t/T_g}
\ee
is only expected to provide a good approximation for long distances which
is indeed the opposite limit that we are studying. That is, we may only
expect (\ref{tg}) to provide a good description of the leading NP
corrections when (\ref{me}) is satisfied and\footnote{In this situation
new NP contributions seem to appear \cite{nos3}.}
$$
{1 \over T_g} >> m \beta_n^2 \,.
$$
Indeed, lattice calculations
\cite{lattice} suggest that the right behavior at long distances is
(\ref{lin}). Therefore, we conclude that (\ref{tg}) can not be applied when the approximation (\ref{ae}) is satisfied which is the situation we are interested in.
\medskip

Let us now summarize our conclusions.

We have calculated the next-to-leading NP corrections to the energy and wavefunctions for heavy quarkonium systems with general quantum numbers $n$ and $l$. We have obtained general analytical expressions. We have also been able to obtain analytical expressions for an arbitrary octet potential. We have also  exactly calculated the error done if we had neglected the octet potential ($N_c \rightarrow \infty$) in our results. We have seen these observables to depend on the three-gluon condensate.

We have reanalyzed the applicability of the VL approach for the lower
levels of Bottomonium and Charmonium. We have studied two possible sets of 
parameters, especially the one 
compatible with the values used in \cite{Ynd1,Ynd2}. We have carefully studied the multipole
expansion (\ref{me}) and the approximation (\ref{ae}), especially the
last one. As we have remarked throughout the paper, this last point is essential in order to distinguish whether a finite $T_g$ is needed. We also believe our
work clarifies whether the $\alpha_s$ perturbative expansion can be applied.

It has been proven that $M_{\Upsilon(1S)}$ lies inside the VL regime since
the NP corrections are under control and very small. It follows that neither
(\ref{tg}) nor potentials models should be applied here. Therefore, this
observable could provide one of the best model independent determination of the $m_b$ mass
where the principal source of error will come from $\Lambda_{QCD}$ and
higher orders in the perturbative expansion which are calculable and do not 
introduce any new parameter. Nevertheless, these corrections could be large. 
Other very striking consequence is that $M_{\Upsilon(1S)}$ should not be used fixing parameters in the potentials models. This would obviously change the value of these parameters. It would be rather interesting to see how important these changes are.  Finally, we have also seen the weak $\mu$ scale
dependence of the result.

For the $n=1$ Charmonium level, using the first set of parameters, the theory 
seems to behave properly and
the VL approach to work but the largeness of $\alpha_s$ makes the results
 less reliable.

For the first set of parameters 
we have also shown that the VL approach is not really applicable for
$n=2$ states and probably both assumptions the multipole expansion (\ref{me}) and the approximation (\ref{ae})
fail for those states and obviously for higher levels.
Therefore, in order to control the NP contributions,
we should look for other approaches, perhaps in the spirit of ref. \cite{DS}, where the NP contributions
are not considered to be corrections. 

The $\Upsilon(1S)$ decay width is very interesting. We have seen that calculating naively we do not
obtain agreement
with the data. The reason is that the new NP contribution blows up
before we can reach the scale used in ref. \cite{Ynd2}.
It has been noted that the $\Upsilon(1S)$ decay width is a two-scale problem and, in
 the spirit of NRQCD, different scales should be used for the different
contributions. We have argued about whether agreement with the data can be obtained. Nevertheless, errors are large owing to the large value of
$\delta_{WF}$ for scales $\mu \sim a_{bb,1}^{-1}$. The solution could be to 
calculate the two loop ${\bar
Q}-Q $ potential which could fix the value of $\delta_{WF}$. This may 
significantly improve the agreement with the experiment. We have also seen that
the multipole expansion (\ref{me})
 and the approximation (\ref{ae}) applies for scales $\mu \sim a_{bb,1}^{-1}$ ((\ref{me}) is especially well satisfied).

We have not performed
the calculation for the fine and hyperfine splitting but we believe that
they will behave in the same way than the $\Upsilon(1S)$ decay width,
being strong scale dependent.

Improvements of our results would come from diminishing errors
in the parameters. It would also be very interesting to have the two loop 
$Q-{\bar Q}$
potential
in order to obtain the next-to-next-to-leading perturbative contribution
to masses and decay widths. It would improve the $m_b$ results and
the
$\Upsilon(1S)$ decay
width. As we said, for the latter it could be essential in order to set the 
right value of $\delta_{WF}$. For the $\eta_c$ and $n=2$ Bottomonium states 
the knowledge of
the next-to-next-to-leading perturbative contribution could eventually
discern whether the $\alpha_s$ expansion works or not. Moreover, a complete  
analysis with the second set of parameters remains to be done for these 
states (the conclusions obtained above could change for this set of 
parameters). This analysis will be only reliable once the above perturbative contributions are calculated.

When available, our results may eventually fix parameters in
NRQCD and HQET. For both of them we can fix $m_b$ and $m_c$ in a model independent way. Once the masses are known we can obtain the HQET parameter ${\bar \Lambda}$ as it has already been done above. We also expect to be able to give QCD rigorous predictions of NRQCD matrix elements related with the $\Upsilon(1S)$ decay width. Moreover, this theoretical framework is able to deal with octet matrix elements in a natural way so further predictions may, in principle, be given.
\medskip

{\bf Acknowledgments}
\medskip

I acknowledge a fellowship from Generalitat of Catalunya.
 Financial support from CICYT, contract AEN95-0590 and financial
support from CIRIT, contract GRQ93-1047 is also acknowledged.
I thank J.I. Latorre for useful discussions. I thank J. Soto for the critical reading of the manuscript and illuminating conversations.

\bigskip

\appendix

\section{Formulas of sec. 2}
\indent

In this section we display analytical expressions and numerical values
for the formulas obtained in sect. 2.

\bea
\nonumber
&&H(n,l)=
{{1024\,{n^8}
   }\over
   {81\,\left(  9\,n^2 - 64 \right) \,
     {{\left( 81\,n^2 - 64 \right) }^3}\,
     {{\left( 81\,n^2 - 256 \right) }^3}}}
\\
&&
\nonumber
\times
\Biggl( 13295844358881280 - 68153404341354496\,{n^2} +
       145600287615221760\,{n^4}
\\
&&
\nonumber
- 168309372752363520\,{n^6} +
       114216987240880128\,{n^8} - 46158344158975776\,{n^{10}}
\\
&&
\nonumber
     +  10789755579716526\,{n^{12}} - 1327743092409993\,{n^{14}} +
       65241222927111\,{n^{16}} +
\\
&&
\nonumber
       l\,\Bigl( 32604849090592768 - 128931229385883648\,{n^2} +
          198763892651851776\,{n^4}
\\
&&
\nonumber
 - 154057684466810880\,{n^6} +
         65418421737648384\,{n^8} - 15363511613472780\,{n^{10}}
\\
&&
\nonumber
 +
         1857043938050595\,{n^{12}} - 88965303991515\,{n^{14}} \Bigr) +
\\
&&
\nonumber
       {l^2}\,\Bigl( 40459485281517568 - 155542385042915328\,{n^2} +
          231200955030306816\,{n^4}
\\
&&
\nonumber
 - 171789196119588864\,{n^6} +
          70180300981986048\,{n^8} - 15974396240527980\,{n^{10}}
\\
&&
\nonumber
 +
        1886699039381100\,{n^{12}} - 88965303991515\,{n^{14}} \Bigr) +
\\
&&
\nonumber
       {l^3}\,\Bigl( 15799724393103360 - 53517313156055040\,{n^2} +
          65200816249896960\,{n^4}
\\
&&
\nonumber
 - 35606723591467008\,{n^6} +
          9549421221866688\,{n^8} - 1223181401792805\,{n^{10}}
\\
&&
 +
          59310202661010\,{n^{12}} \Bigr)  +
\\
&&
\nonumber
       {l^4}\,\Bigl( 8125992224686080 - 27496161183006720\,{n^2} +
          33417136857415680\,{n^4}
\\
&&
\nonumber
 - 18162612510511104\,{n^6} +
          4838867443911744\,{n^8} - 615121070102415\,{n^{10}}
\\
&&
\nonumber
 +
          29655101330505\,{n^{12}} \Bigr)  +
\\
&&
\nonumber
       {l^5}\,\Bigl( 271356033761280 - 885005525975040\,{n^2} +
          980074478960640\,{n^4}
\\
&&
\nonumber
 - 431100857733120\,{n^6} +
          76988199574080\,{n^8} - 4236443047215\,{n^{10}} \Bigr)  +
\\
&&
\nonumber
       {l^6}\,\Bigl( 90452011253760 - 295001841991680\,{n^2} +
           326691492986880\,{n^4}
\\
&&
\nonumber
- 143700285911040\,{n^6} +
          25662733191360\,{n^8} - 1412147682405\,{n^{10}} \Bigr)
\Biggr)\,,
\eea

\be
H(1,0)= {{141912051712}\over {844421875}}\,, \quad
H(2,0)= {{484859657191424}\over {2040039729}}\,,
\ee
$$
H(2,1)= {{102150951135870976}\over {765014898375}}\,, \quad
H(3,0)= {{1299369918513056329728}\over {79496721732575}}\,,
$$
$$
H(3,1)= {{6471153009519628976971776}\over {529050683130286625}} \,,
\quad H(3,2)= {{2239521974640025214976}\over {397483608662875}} \,.
$$

\bea
\nonumber
&&W(n)=
{512\,{n^{10}}
\over
    {81\,{{\left( 9\,n^2 - 64 \right) }^2}\,
      \left( 81\,n^2 - 1024 \right) \,
      {{\left( 81\,n^2 - 64 \right) }^4}\,
      {{\left( 81\,n^2 - 256 \right) }^4} }}
\\
&&
\nonumber
\times
\biggl( -447359221655207230383849472 +
        3163532546828444207717285888\,{n^2}
\\
&&
\nonumber
 -        9812275706440214903775035392\,{n^4} +
        17549420408786506395496218624\,{n^6}
\\
&&
\nonumber
 -
        20019188843798683542274179072\,{n^8} +
        15231566821196295158466871296\,{n^{10}}
\\
&&
\nonumber
-       7874351861115834918827458560\,{n^{12}} +
        2775326655204987593042165760\,{n^{14}}
\\
&&
\nonumber
 -
        660250519084024545000996864\,{n^{16}} +
        103213969217030305334486016\,{n^{18}}
\\
&&
\nonumber
 -
        10045634813323035262758792\,{n^{20}} +
        545473369543317738837741\,{n^{22}}
\\
&&
 -
        12459805959909788031573\,{n^{24}} \biggr) \,,
\eea

$$
W(1)= -{{1670626488940208128}\over {485563688671875}}\,, \quad
W(2)= -{{525333494026541203456}\over {26299512173025}}\,,
$$
\be
W(3)= -{{6910180342493957842421202407424}\over {2280028384216811071375}}
\,. \ee
\bigskip


\section{Octet potential}
\indent

In this appendix we briefly discuss the role played by the octet coulomb
potential and the error done neglecting it.

Frequently the octet potential is neglected in the octet propagator when
the gluon correlation time is taken into account \cite{Giac1,SY} (see
\cite{Dosch} where the octet potential is taken into account being solved
the differential equations numerically). Therefore, we find pretty
interesting to consider the case for a general octet potential and take
the limit $N_c \rightarrow \infty$ in our results. Thus, we recalculate
$H$ and $W$ with ${1 \over 8} = {1 \over N_c^2-1} \equiv r$ for a
general $r$. That is,
$$
V_8={1 \over 2N_c} {{\tilde \alpha_s} \over r} \rightarrow {1 \over
N_c^2-1}{ C_F {\tilde \alpha_s} \over r}=-rV_s\,,
$$
$$
H(n,l) \rightarrow H_{oct}(n,l,r)\,,
$$
\be
W(n) \rightarrow W_{oct}(n,r)\,.
\ee
We obtain for $r=0$ ($N_c \rightarrow \infty$)
$$
H_{oct}(1,0,0)=198       \sim  1.18 H(1,0)\,, \quad
H_{oct}(2,0,0)=312576    \sim  1.32 H(2,0)\,,
$$
$$
H_{oct}(2,1,0)=183040    \sim  1.37 H(2,1)\,, \quad
H_{oct}(3,0,0)=22609206  \sim  1.38 H(3,0)\,,
$$
$$
H_{oct}(3,1,0)=17058600  \sim  1.39 H(3,1)\,, \quad
H_{oct}(3,2,0)=7846956   \sim  1.39 H(3,2)\,,
$$
$$
W_{oct}(1,0)=-{20548 \over 5}       \sim  1.19 W(1)\,, \quad
W_{oct}(2,0)=-{133338112 \over 5}   \sim  1.34 W(2) \,,
$$
\be
W_{oct}(3,0)=-{21133755198 \over 5} \sim  1.39 W(3) \,.
\ee

We find they are always larger than the values obtained with $N_c=3$. We
also see that the error is large though under control. Finally, we have
also been able to obtain analytical expressions for $W$ and $H$ (with
$l=0$) for a general $r$. They read

\bea
\label{Woct}
\nonumber
&&W_{oct}(n,r) =
 288\,n^{10} \,\Bigl( {{3\,\left( 1 - {n^{2}} \right) }\over
       {2\,{{\left( 1 + r \right) }^4}}} +
     {{17\,\left( 1 - {n^{2}} \right) }\over
       {3\,{{\left( 1 + r \right) }^3}}} +
     {{\left( 898 - 873\,{n^2} - 25\,{n^4} \right) }\over
       {72\,{{\left( 1 + r \right) }^2}}}
\\
&&
\nonumber
 +
     {{\left( 17976 - 17741\,{n^2} - 235\,{n^4} \right) }\over
       {864\,\left( 1 + r \right) }} +
     {{{n}\,\left( 120 - 274\,n + 225\,{n^2} - 85\,{n^3} +
           15\,{n^4} - {n^5} \right) }\over
       {17280\,\left( -4 + n + n\,r \right) }}
\\
&&
\nonumber
 +
     {{{{\left( -3 + n \right) }^2}\,
     \left( -40 + 94\,n - 77\,{n^2} + 26\,{n^3} - 3\,{n^4} \right)
}\over
       {1440\,{{\left( -3 + n + n\,r \right) }^2}}}
\\
&&
\nonumber
 +
     {{\left( 2520 - 9186\,n + 13531\,{n^2} - 10240\,{n^3} +
           4160\,{n^4} - 854\,{n^5} + 69\,{n^6} \right) }\over
       {4320\,\left( -3 + n + n\,r \right) }}
\\
&&
\nonumber
 +
     {{{{\left( -2 + n \right) }^4}\,
         \left( -3 + 4\,n - {n^2} \right) }\over
       {24\,{{\left( -2 + n + n\,r \right) }^4}}}
 +    {{{{\left( 2 - n \right) }^3}\,
         \left( 108 - 75\,n - 56\,{n^2} + 23\,{n^3} \right) }\over
       {432\,{{\left( -2 + n + n\,r \right) }^3}}}
\\
&&
\nonumber
 +
     {{{{\left( -2 + n \right) }^2}\,
         \left( -360 - 4743\,n + 8026\,{n^2} - 3317\,{n^3} +
           394\,{n^4} \right) }\over
       {4320\,{{\left( -2 + n + n\,r \right) }^2}}}
\\
&&
\nonumber
 +    {{\left( -10080 + 99624\,n - 142806\,{n^2} + 42615\,{n^3} +
           22325\,{n^4} - 13479\,{n^5} + 1801\,{n^6} \right) }\over
       {17280\,\left( -2 + n + n\,r \right) }}
\\
&&
\nonumber
 +
     {{2\,\left( 2 - n \right) \,{{\left( -1 + n \right) }^4}\,
         {n}}\over {3\,{{\left( -1 + n + n\,r \right) }^4}}} +
     {{{{\left( -1 + n \right) }^3}\,{n}\,
         \left( 126 - 53\,n - 5\,{n^2} \right) }\over
       {27\,{{\left( -1 + n + n\,r \right) }^3}}}
\\
&&
\nonumber
 +
     {{{{\left( -1 + n \right) }^2}\,{n}\,
         \left( 8070 - 1103\,n - 1612\,{n^2} + 73\,{n^3} \right) }\over
       {864\,{{\left( -1 + n + n\,r \right) }^2}}}
\\
&&
\nonumber
 +    {{{n}\,\left( -121380 + 87068\,n + 87765\,{n^2} - 60855\,{n^3} +
           9615\,{n^4} - 2213\,{n^5} \right) }\over
       {8640\,\left( -1 + n + n\,r \right) }}
\\
&&
\nonumber
 -
 {{2\,{n}\,{{\left( 1 + n \right) }^4}\,\left( 2 + n \right) }\over
       {3\,{{\left( 1 + n + n\,r \right) }^4}}} +
     {{{n}\,{{\left( 1 + n \right) }^3}\,
         \left( -126 - 53\,n + 5\,{n^2} \right) }\over
       {27\,{{\left( 1 + n + n\,r \right) }^3}}}
\\
&&
\nonumber
 +
     {{{n}\,{{\left( 1 + n \right) }^2}\,
         \left( -8070 - 1103\,n + 1612\,{n^2} + 73\,{n^3} \right) }\over
       {864\,{{\left( 1 + n + n\,r \right) }^2}}}
\\
&&
\nonumber
 +    {{{n}\,\left( -121380 - 87068\,n + 87765\,{n^2} + 60855\,{n^3} +
           9615\,{n^4} + 2213\,{n^5} \right) }\over
       {8640\,\left( 1 + n + n\,r \right) }}
\\
&&
\nonumber
 -
     {{{{\left( 2 + n \right) }^4}\,
         \left( 3 + 4\,n + {n^2} \right) }\over
       {24\,{{\left( 2 + n + n\,r \right) }^4}}} +
     {{{{\left( 2 + n \right) }^3}\,
         \left( -108 - 75\,n + 56\,{n^2} + 23\,{n^3} \right) }\over
       {432\,{{\left( 2 + n + n\,r \right) }^3}}}
\\
&&
\nonumber
+
     {{{{\left( 2 + n \right) }^2}\,
         \left( -360 + 4743\,n + 8026\,{n^2} + 3317\,{n^3} +
           394\,{n^4} \right) }\over
       {4320\,{{\left( 2 + n + n\,r \right) }^2}}}
\\
&&
\nonumber
 +    {{\left( 10080 + 99624\,n + 142806\,{n^2} + 42615\,{n^3} -
           22325\,{n^4} - 13479\,{n^5} - 1801\,{n^6} \right) }\over
       {17280\,\left( 2 + n + n\,r \right) }}
\\
&&
\nonumber
 -
     {{{{\left( 3 + n \right) }^2}\,
      \left( 40 + 94\,n + 77\,{n^2} + 26\,{n^3} + 3\,{n^4} \right)
}\over
      {1440\,{{\left( 3 + n + n\,r \right) }^2}}}
\\
&&
\nonumber
 -    {{\left( 2520 + 9186\,n + 13531\,{n^2} + 10240\,{n^3} +
           4160\,{n^4} + 854\,{n^5} + 69\,{n^6} \right) }\over
       {4320\,\left( 3 + n + n\,r \right) }}
\\
&&
 +
     {{{n}\,\left( 120 + 274\,n + 225\,{n^2} + 85\,{n^3} + 15\,{n^4} +
     {n^5} \right) }\over {17280\,\left( 4 + n + n\,r \right) }}
\Bigr) \,,
\eea
\medskip

\bea
\label{Eoct}
\nonumber
&&H_{oct}(n,0,r)=
n^8 \Bigl( {{72\,\left( -1 + {n^{2}} \right) }\over
     {{{\left( 1 + r \right) }^3}}} +
   {{228\,\left( -1 + {n^{2}} \right) }\over
     {{{\left( 1 + r \right) }^2}}} +
{{2\,\left( -214 + 229\,{n^2} - 15\,{n^4} \right) }\over {1 + r}}
\\
&&
+   {{{{\left( -1 + n \right) }^2}\,
       \left( -24 + 26\,n - 9\,{n^2} + {n^3} \right) }\over
     {2\,\left( -3 + n + n\,r \right) }} +
   {{2\,{{\left( -2 + n \right) }^3}\,
       \left( 3 - 4\,n + {n^2} \right) }\over
     {{{\left( -2 + n + n\,r \right) }^3}}}
\\
&&
\nonumber
 +
   {{{{\left( -2 + n \right) }^2}\,
       \left( 3 + 20\,n - 31\,{n^2} + 8\,{n^3} \right) }\over
     {{{\left( -2 + n + n\,r \right) }^2}}} +
   {{\left( 12 - 112\,n + 117\,{n^2} + 18\,{n^3} - 45\,{n^4}
      +         10\,{n^5} \right) }\over {-2 + n + n\,r}}
\\
&&
\nonumber
 +
   {{32\,\left( -2 + n \right) \,{{\left( -1 + n \right) }^3}\,{n}}\over
     {{{\left( -1 + n + n\,r \right) }^3}}} +
   {{8\,{{\left( -1 + n \right) }^2}\,{n}\,
       \left( -22 + n + 5\,{n^2} \right) }\over
     {{{\left( -1 + n + n\,r \right) }^2}}} 
\\
&&
\nonumber
+ {{{n}\,\left( 578 - 303\,n - 539\,{n^2} + 255\,{n^3} + 9\,{n^4}
\right)
        }\over {2\,\left( -1 + n + n\,r \right) }} +
  {{32\,{n}\,{{\left( 1 + n \right) }^3}\,\left( 2 + n \right) }\over
     {{{\left( 1 + n + n\,r \right) }^3}}}
\\
&&
\nonumber
 +
   {{8\,{n}\,{{\left( 1 + n \right) }^2}\,
       \left( 22 + n - 5\,{n^2} \right) }\over
     {{{\left( 1 + n + n\,r \right) }^2}}} +
{{{n}\,\left( 578 + 303\,n - 539\,{n^2} - 255\,{n^3} + 9\,{n^4}
\right)
          }\over {2\,\left( 1 + n + n\,r \right) }}
\\
&&
\nonumber
 +
   {{2\,{{\left( 2 + n \right) }^3}\,
       \left( 3 + 4\,n + {n^2} \right) }\over
     {{{\left( 2 + n + n\,r \right) }^3}}} +
   {{{{\left( 2 + n \right) }^2}\,
       \left( 3 - 20\,n - 31\,{n^2} - 8\,{n^3} \right) }\over
     {{{\left( 2 + n + n\,r \right) }^2}}} 
\\
&&
\nonumber
+   {{\left( -12 - 112\,n - 117\,{n^2} + 18\,{n^3} + 45\,{n^4} +
         10\,{n^5} \right) }\over {2 + n + n\,r}} +
   {{{{\left( 1 + n \right) }^2}\,
       \left( 24 + 26\,n + 9\,{n^2} + {n^3} \right) }\over
     {2\,\left( 3 + n + n\,r \right) }} \Bigl) \,.
\eea
(\ref{Woct}) and (\ref{Eoct}) do not have a unique limit when $n
\rightarrow integer$, $r \rightarrow 0$. The proper order is

$$\lim_{r
\rightarrow 0} \lim_{{\scriptscriptstyle{n \rightarrow integer}}}\,.$$

\bigskip


\end{document}